\documentclass[sigconf]{acmart}
\AtBeginDocument{%
  \providecommand\BibTeX{{%
    \normalfont B\kern-0.5em{\scshape i\kern-0.25em b}\kern-0.8em\TeX}}}

\setcopyright{acmcopyright}
\copyrightyear{2021}
\acmYear{2021}

\acmConference[SIGIR '21]{SIGIR '21: ACM Special Interest Group on Information Retrieval}{July, 2021}{Montreal, Canada}

\begin{document}

\title{Non-Linear Multiple Field Interactions Neural Document Ranking}


\author{Kentaro Takeguchi}
\email{gm19099@bristol.ac.uk}
\affiliation{%
  \institution{University of Bristol}
  \country{United Kingdom}
}

\author{Niall Twomey}
\email{niall-twomey@cookpad.com}
\affiliation{%
  \institution{Cookpad Ltd}
  \country{United Kingdom}}

\author{Luis M Vaquero}
\email{luis.vaquero@bristol.ac.uk}
\affiliation{%
  \institution{University of Bristol}
  \country{United Kingdom}
}

\renewcommand{\shortauthors}{Takeguchi, et al.}

\begin{abstract}
Ranking tasks are usually based on the text of the main body of the page and the actions (clicks) of users on the page. There are other elements that could be leveraged to better contextualise the ranking experience (e.g. text in other fields, query made by the user, images, etc). We present one of the first in-depth analyses of field interaction for multiple field ranking in two separate datasets. While some works have taken advantage of full document structure, some aspects remain unexplored. In this work we build on previous analyses to show how query-field interactions, non-linear field interactions, and the architecture of the underlying neural model affect performance.
\end{abstract}

\ccsdesc[500]{Applied computing~Document management and text processing}
\ccsdesc[500]{Information systems~Learning to rank}

\keywords{recipes, neural networks, field interactions, query-field, first-order, document ranking}

\maketitle

\section{Introduction}

Modern online documents consists of a number of fields, such as title, body, the anchor text from incoming hyperlinks, or the query text for which the document has been previously viewed.

However, recent efforts have suggested that field dependencies are not critical for some search applications\cite{mitra2019incorporating,sarvi2020comparison}.

This is a counter-intuitive result as one may think multiple fields associated with each document may contain complementary information; this, in turn, can improve the performance of the ranking task. 

This intuition was exploited by traditional information retrieval techniques where field interactions are explicitly considered and deemed to be important~\cite{Robertson2004}. A number of techniques has been explored to represent these interactions and their impact in information retrieval~\cite{piwowarski2003machine, svore2009machine, kim09prob, kim2012field}.

Until recently, deep neural ranking models (NRMs) tended to consider a single source of document description, such as document title~\cite{huang2013learning} or body text~\cite{Dehghani2017nrmweak}. Interactions between fields have recently started to be considered in NRMs. 

Field relevance modelling is usually conceived in two different ways: 1) from fields represented independently and then combined to create a weighted mode~\cite{Binsheng18field-based-relevance, DBLP:journals/corr/abs-1711-09174} or 2) directly from the entire document and using a relevance model to score fields~\cite{Binsheng18field-based-relevance}.


While~\cite{DBLP:journals/corr/abs-1711-09174} explicitly learn query-field interactions, no prior works have not explicitly explored the nature of field-to-field interactions to better understand their inter-relations. Simple combinations of relevance on each field~\cite{Binsheng18field-based-relevance} or field representation concatenations~\cite{DBLP:journals/corr/abs-1711-09174} have been used. Like in attention models for natural language processing, it may be that some interactions between fields are non linear in nature. The order of field-to-field non linearities and their impact on ranking performance is unexplored.

We evaluate our models in the context of web search, using
the queries sampled from the Cookpad's`\footnote{http://www.cookpad.com} search logs. We study multiple field combinations to understand if more complex non-linear combinations can have an impact on retrieval performance.

In order to assess the robustness of our findings, this work also explores the impact of the architectural model employed for the information retrieval task.

In this work, we study the following research hypotheses:
\begin{itemize}
    \item \textbf{H1} Non-linearities in field-to-field interactions have an impact in ranking performance.
    \item \textbf{H2} Specific field-to-field interactions  affect performance (beyond query-field interactions).
    \item \textbf{H3} The importance of field interactions are dependent on the neural architecture employed for the analysis.
\end{itemize}

Our experiments validate all these hypotheses, and investigate
the effectiveness of our overall exploration of field interactions.



\section{Related Work}

\subsection{Retrieval with Multiple Fields}


Classic works already relayed on the usage of information from multiple fields in a document~\cite{robertson1995okapi}. Robertson et al.~\cite{Robertson2004} extended the original BM25 model to create the BM25F model, which combines frequency information across fields on a per-term basis and then computes a retrieval score using a balanced approach. 

Other approaches built on this idea without resorting to a linear combination of per-field scores: like, for instance, Bayesian networks~\cite{piwowarski2003machine}, LambdaBM25~\cite{svore2009machine} (based on LambdaRank~\cite{burges2007learning}), language modeling framework~\cite{Ogilvie03known}, probabilistic models~\cite{kim09prob}, or feedback weighted field relevance~\cite{kim2012field}.

\subsection{Neural Networks for Ranking}

Mitra et al. have shown that no significant loss is observed in  models that incorporate the query term independence assumption (order of words does not matter and relevance is only measured if a word is in the query) in web search \cite{mitra2019incorporating}. 

State-of-the-art BERT-based models are mediocre in product search~\cite{sarvi2020comparison}. Together, these works suggest that capturing inter-term dependencies is not critical in some search applications, even though it is a central concern in question answering.

While there are many works focusing on the application of neural models to information retrieval~\cite{MitraandCraswellNIR18}, but most of them treat each document as a single instance of text (i.e., single field). 

However, documents often include information in a semi-structured format and multiple fields.  A few studies have discussed how to use evidence from structure to improve the performance of information retrieval systems. 

Wilkinson proposed several heuristic methods of combining section-level and document-level evidence, such as taking the maximum section score or taking a weighted sum of section scores \cite{wilkinson1994effective}.

NRM-F,  proposed by Zamani et al. \cite{DBLP:journals/corr/abs-1711-09174}, is the only paper that discusses how neural models can deal with multiple document fields from an architectural perspective. The authors say that it is better for the ranker to score the whole document jointly, rather than generate a per-field score and aggregate. 

NRM-F formulates the document representation learning function $\Phi_D$ as follows:
$$
\Phi_{D}\left(F_{d}\right)=\Lambda_{D}\left(\Phi_{F_{1}}\left(F_{1}\right), \Phi_{F_{2}}\left(F_{2}\right), \cdots, \Phi_{F_{k}}\left(F_{k}\right)\right)
$$
where $\Phi_{F_{i}}$ denotes the mapping function for the field $F_i$ and $\Lambda_D$ aggregates representations learned for all the fields. $\Lambda_{D}$ simply concatenates the input vectors to be served in the matching function. Then, a stack of fully-connected layers outputs the final retrieval score.

In NRM-F, both query text and text fields are represented using a character $n$-gram hashing vector as in \cite{huang2013learning}. Then, a convolution layer is employed to capture the dependency between terms.

This model explicitly learns query-field interactions, but it does not distinctly consider field-to-field interactions. Importantly, the effect of non-linear interactions between fields is also not taken into account. NRM-F also exclusively focuses on query-field interactions (scoring the whole document jointly), but there may be other important field interactions to consider. 

\cite{Binsheng18field-based-relevance} also focus on query-to-field interactions and assume there are just linear relationships between relevance models induced from each fields.

When designing a ranking model, several architecture decisions need to be made, such as representation-based vs. interaction-based, which field interactions to learn, how to aggregate scores, etc. These fundamental design decisions were not clearly justified in prior efforts and the impact of the chosen architectural model on ranking remains unclear.

There are several potential models that have been used to assess the effectiveness of field-interactions.Factorization  Machine  (FM)  is  a  widely  used  supervised learning approach by effectively modeling of feature interactions. In FM, unseen feature interactions can be learned from other pairs. Field-weighted Factorization Machine (FwFM) are state-of-the-art among the shallow models for click-through-rate prediction~\cite{pan2018field}.

\section{Methodology}

\subsection{Datasets}

\subsubsection{MS MARCO Dataset}
Microsoft has released a large web search dataset called MS MARCO \footnote{MS MARCO \url{https://microsoft.github.io/msmarco/}}. The documents consist of multiple fields. 

\subsubsection{Recipe Search Dataset}

Cookpad\footnote{Cookpad \url{http://www.cookpad.com}} is the number one Japanese online recipe community platform. Users can publish and search for recipes on the platform. The size of the dataset is shown in Table \ref{tab:basic_statistics_of_the_dataset}. The dataset consists of two subsets: master recipe data and search logs.

\begin{table}
    \centering
    \begin{tabular}{l|l}
        \hline
        \#queries & 2,733,549 \\
        \#unique queries & 94,993 \\
        \#unique presented recipes & 301,078 \\
        \hline
    \end{tabular}
    \caption{Basic Statistics of the Cookpad dataset}
    \label{tab:basic_statistics_of_the_dataset}
\end{table}

Recipes are structured data as shown in Table \ref{tab:schema_of_recipe_data}. The description is a free text field; some recipes have a surprisingly long description while there are recipes with no description. The ingredients are an unsorted set of entities. 

\begin{table}
    \centering
    \begin{tabular}{l|l|l}
        \hline
        Field & Type & Example \\
        \hline
        recipe\_id & Integer & 1 \\
        title & String & Honey garlic chicken thighs \\
        description & String & This recipe has always been my favorite \\
        ingredients & String set & chicken, salt, crushed red chilli, ... \\
        country & String & GB \\
        \hline
    \end{tabular}
    \caption{The schema of recipe data}
    \label{tab:schema_of_recipe_data}
\end{table}

\begin{figure}
    \centering
    \includegraphics[width=0.5\textwidth]{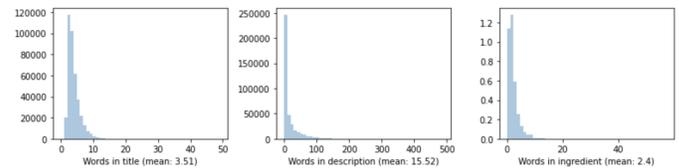}
    \caption{Number of words in text fields in the Cookpad data}
    \label{fig:words_in_fields}
\end{figure}

Figure \ref{fig:words_in_fields} shows the distribution of the number of words in each field; text fields are generally short in length, making it difficult to capture text significance using term frequency-based methods.

The second subset of data is search logs: event log created when a user clicks a recipe in the search results. The attributes of each event are listed in Table \ref{tab:schema_of_search_log}. \textit{fetched\_recipe\_id} indicates what recipes were retrieved against the query and \textit{position} shows the clicked recipe position in the list.

\begin{table}
    \centering
    \begin{tabular}{l|l|l}
        \hline
        Field & Type & Example \\
        \hline
        session\_id & Integer & 1 \\
        query & String & hot dessert \\
        page & Integer & 1 \\
        recipe\_id & Integer & 1 \\
        position & Integer & 1 \\
        fetched\_recipe\_ids & String & 1,2,3,4,5,6 \\
        total\_hits & Integer & 256 \\
        \hline
    \end{tabular}
    \caption{The schema of search log in Cookpad}
    \label{tab:schema_of_search_log}
\end{table}

\begin{figure}
    \centering
    \includegraphics[width=0.3\textwidth]{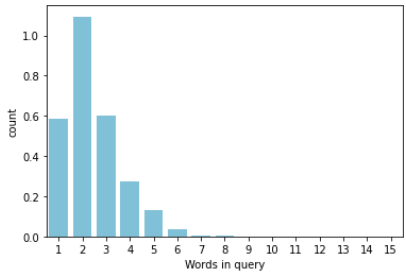}
    \caption{The number of words per query in the Cookpad data}
    \label{fig:words_in_query}
\end{figure}

Figure \ref{fig:words_in_query} shows the distribution of words in a query. Most queries contain no more than three words.

\subsection{Data Processing and Modeling}
\label{subsec:data processing and modeling}

We selected five fields: query, title, description, ingredients, and country for training. The text embedding is shared across text fields. This data is concatenated with search logs.

Search logs are aggregated by session ID and query, and the result list is trimmed at the clicked position. Labels are then assigned by referring to the position. 

We employ pairwise cross-entropy loss.

Regarding the text representation, we obtain fix-sized vectors under the assumption that terms are almost independent. In this type of application, it is not uncommon to assume term independence. Amazon's experiment showed that taking the average of term vectors performed similar or slightly better than recurrent units with significantly less training time \cite{DBLP:journals/corr/abs-1907-00937}. The countries are treated as a category and embedded into a latent space.

In order to tests if field interactions affect ranking performance in an architecture-dependent manner, we focus on two architectural models: NRM-F~\cite{DBLP:journals/corr/abs-1907-00937} and FwFM~\cite{pan2018field} to examine how the choice of architecture affects effectiveness. The table summarises the differences between those two models.

\begin{table}[H]
    \centering
    \begin{tabular}{l|l|l}
        \hline
        & NRM-F & FwFM \\
        \hline
        First-order features & Not used & Used \\
        Interaction selection & Query-field & All \\
        Interaction representation & Hadamard product & Dot product \\
        Interaction aggregation & Concatenation & Summation \\
        \hline
    \end{tabular}
    \caption{A Comparison of Different Architectures}
    \label{tab:differences_in_models}
\end{table}

We employ Normalised Discounted Cumulative Gain (NDCG) to evaluate models, with a cut-off of 20~\footnote{this is the number of recipes served per page at Cookpad}. 

The entire dataset is divided into 10 sets by timestamp to obtain a sufficient number of individual datasets to evaluate the statistical significance of the obtained results. Each dataset is further divided by timestamp, with the first 75\% used for training and the remaining 25\% for validation. 

\section{Experiments}
All the experiments presented in this section are available for reproducibility\footnote{\url{https://github.com/rejasupotaro/master-thesis}}.

\subsection{Impact of Field Interactions}

First, we determine whether field interactions have any effect in ranking for our datasets. In order to accomplish this we use:

\begin{itemize}
    \item \textbf{A representation-based model (No feature interaction)} As a no interaction learning model, we employ a simple representation-based model that consists of two component: query encoder and recipe encoder. Both encoders transform entities into vectors and their cosine similarity is computed at the last layer.
    \item \textbf{An implicit interaction-based model}: The naive interaction-based model simply concatenates all features at the first layer, then the output is fed to several fully-connected layers.
    \item \textbf{An NRM-F-based model)}: This model is based on NRM-F, but the text representation was simplified in accordance with the dataset as mentioned above.
\end{itemize}

\begin{table}
    \centering
    \begin{tabular}{l|c}
        \hline
        Model & NDCG@20 \\
        \hline
        No interaction learning & 0.6376 \\
        Implicit interaction learning & 0.6429 \\
        Explicit interaction learning & \textbf{0.6483} \\
        \hline
    \end{tabular}
    \caption{Performance comparison of interaction learning}
    \label{tab:rq1_result}
\end{table}

Table \ref{tab:rq1_result} shows that the interaction-based models outperformed the representation-based model in average performance. Table \ref{tab:rq1_p_values_on_pairs} shows that no statistical significance is observed when performing Tukey's multiple comparison test to determine if there are statistically differences between those models. 

\begin{figure}
    \centering
    \includegraphics[width=0.3\textwidth]{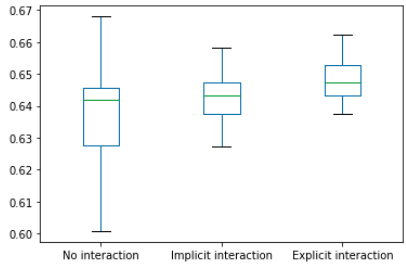}
    \caption{Boxplot showing NDCG scores on whether feature interactions have an impact on our dataset}
    \label{fig:rq1_boxplot}
\end{figure}

Figure \ref{fig:rq1_boxplot} is the boxplot showing the performance of each model. The performance of the representation-based model fluctuates.

\begin{table}
    \centering
    \begin{tabular}{l|c}
        \hline
        Pair & p-value \\
        \hline
        No interaction - Implicit interaction & 0.619 \\
        No interaction - Explicit interaction & 0.159 \\
        Implicit interaction - Explicit interaction & 0.616 \\
        \hline
    \end{tabular}
    \caption{P-values of Tukey's test on pairs}
    \label{tab:rq1_p_values_on_pairs}
\end{table}

\subsection{Query-to-field vs Field-to-field Interactions}
\label{subsec:q2ff2f}

Query-field interactions are considered to be important in document ranking, whereas recommendation models do not distinguish between context and item features. 

Inspired by recommendation models, which do not usually distinguish between a query and item features, we gauge the effects of adding all field interactions vs focusing on query-field interactions only. 

The purpose of this experiment is to investigate whether limiting interactions to just query-field helps to improve performance. To do so, we train two models with different feature interactions:

\begin{itemize}
    \item \textbf{NRM-F (query-field)}: Consider query-field interactions (as the original implementation).
    \item \textbf{NRM-F (all)}: Consider all feature interactions without distinguishing between query and fields.
    \item \textbf{FwFM (all)} : Consider all feature interactions without distinguishing between query and fields (as the original implementation).
    \item \textbf{FwFM (query-field)}: Consider query-field interactions.
\end{itemize}

Table \ref{tab:rq2_result} shows the performance of the above mentioned models. Interestingly, the models that learned query-field interactions outperformed models trained using all interactions in both NRM-F and FwFM. Table \ref{tab:rq2_p_values} shows these differences are statistically significant.

The experiment also has shown that recommendation models could be used for ranking tasks as it is since FwFM outperformed the simplified NRM-F. Besides, we can further improve performance by incorporating the properties of information retrieval into recommendation models.

\begin{table}
    \centering
    \begin{tabular}{l|c}
        \hline
        Model & NDCG@20 \\
        \hline
        NRM-F-based model (query-field interactions) & \textbf{0.6483} \\
        NRM-F-based model (all interactions) & 
        \textbf{0.6403} \\
        FM-based model (query-field interactions) & \textbf{0.6674} \\
        FM-based model (all interactions) &  \textbf{0.6616} \\
        \hline
    \end{tabular}
    \caption{Performance comparison of models with different interactions}
    \label{tab:rq2_result}
\end{table}

\begin{table}
    \centering
    \begin{tabular}{l|c}
        \hline
        Pair & p-value \\
        \hline
        NRM-F (query-field) - NRM-F (all) & 0.0031 \\
        FwFM (query-field) - FwFM (all) & 0.001 \\
        \hline
    \end{tabular}
    \caption{P-values of paired t-tests on each method}
    \label{tab:rq2_p_values}
\end{table}


\subsection{Importance of Interactions beyond Query-Fields}

Subsection~\ref{subsec:q2ff2f} suggests some naturally leads to wondering whether other field interactions beyond query-field interactions can be identified.

In this set of experiments, we employed the FwFM model, trained using first- and second-order interactions (5 features + 10 feature interactions in total). FMs compute the scores for each field independently and sum them up to produce the final score. We trained the model regularly and extracted the individual feature scores on validation data as shown in the following code snippet.

\begin{verbatim}
class FwFM(BaseModel):
    def build(self):
        ...
        x = tf.concat([first_order_features, feature_interactions], axis=1)
        # It is individually computed scores.
        scores = tf.keras.Model(inputs=inputs, outputs=x)
        # Sum up the computed scores, which will be the final score.
        x = tf.keras.backend.sum(x, axis=1, keepdims=True)
        output = layers.Activation('sigmoid', name='label')(x)
        final_score = tf.keras.Model(inputs=inputs, outputs=output, name=self.name)
        # `scores` model is used to extract individual 
        # scores.
        return final_score, scores
\end{verbatim}

Features are sorted by correlation to the label building on the assumption that the correlation should be a proxy indicator for field importance since the sum of individual scores will be the final score. Then, we compare the performance of these three models.

\begin{figure*}
    \centering
    \includegraphics[width=0.99\textwidth]{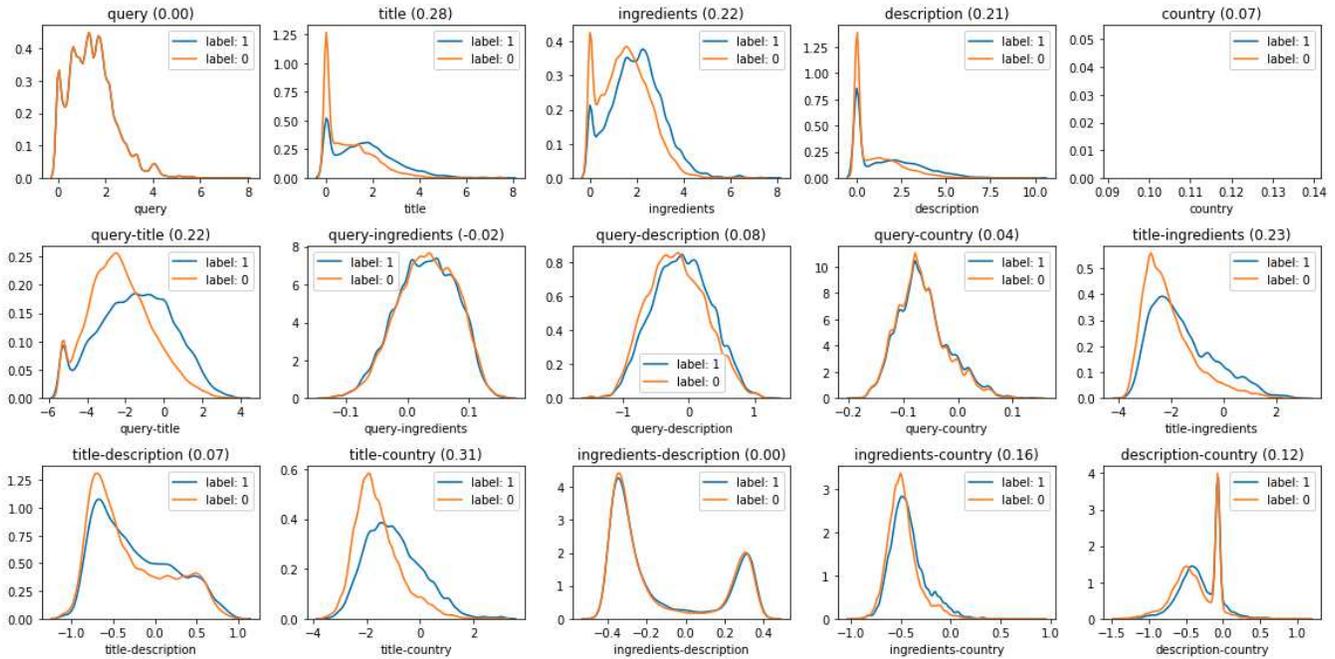}
    \caption{Distribution of outputs of the interaction layer. The label represents 1: clicked and 0: not clicked. The x-axis shows the actual value, and the y-axis is the probability density. The numbers in parentheses indicate the correlation with the label.}
    \label{fig:rq4_distplot}
\end{figure*}

\begin{figure}
    \centering
    \includegraphics[width=0.5\textwidth]{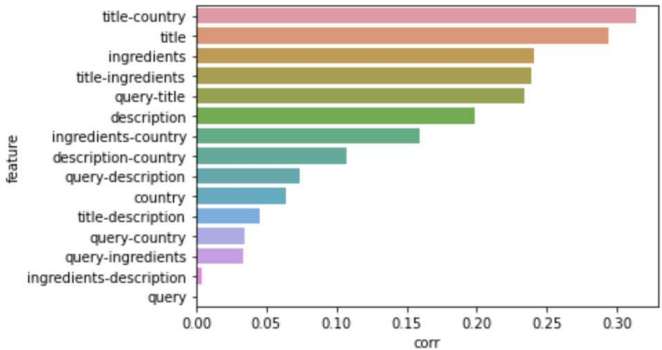}
    \caption{The correlation to the label of each feature. }
    \label{fig:rq4_corr}
\end{figure}

Figure \ref{fig:rq4_distplot} shows the distributions of the activation of fields. The shape of the distribution of query-title is different to that of other fields. Clicked and not clicked distributions coincide: their correlation to the label is zero (see Figure \ref{fig:rq4_corr}). 

These results suggest that the model cannot guess which recipe is more likely to be engaged with just by looking at a query. Also, correlation seems to be associated to the importance of the features. 

\begin{table}
    \centering
    \begin{tabular}{l|c}
        \hline
        Model & NDCG@20 \\
        \hline
        FwFM (all) & 0.662 \\
        FwFM (query-field) & \textbf{0.667} \\
        FwFM (selected) & 0.665 \\
        \hline
    \end{tabular}
    \caption{Performance comparison of models with different interactions}
    \label{tab:rq4_result}
\end{table}

However, table \ref{tab:rq4_result} shows the performance of the FwFMs with different features: the model with selected features did not outperform the model with query-field interactions.

\subsection{Non-Linear Field Interactions}

The original implementation of NRM-F does not use first-order field interactions. In this section, we explore the impact of first-order field interactions in performance. For this set of experiments, we use the following fields: query, title, description, ingredient, and country. We define this non-linear, second-order interactions to use vary by model as follows:

\begin{itemize}
    \item \textbf{NRM-F (2nd)}: Use second-order query-field interactions only (as the original implementation).
    \item \textbf{NRM-F (1st + 2nd)}: Use first-order features along with second-order query-field interactions.
    \item \textbf{FwFM (1st + 2nd)} : Use first- and second-order interactions (as the original implementation).
    \item \textbf{FwFM (2nd)}: Use second-order interactions only.
\end{itemize}

\begin{figure}
    \centering
    \includegraphics[width=0.4\textwidth]{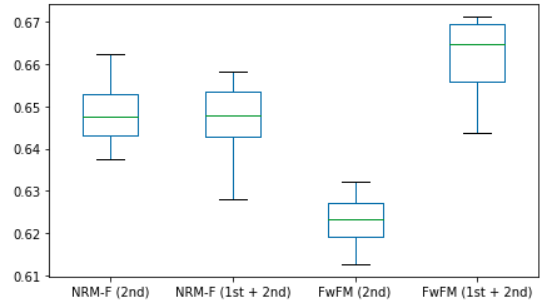}
    \caption{Boxplot showing NDCG scores}
    \label{fig:rq3_boxplot}
\end{figure}

Figure \ref{fig:rq3_boxplot} shows the performance of the models above. It can be seen that there is no difference between NRM-F (2nd) and NRM-F (1st + 2nd) while FwFM (1st + 2nd) significantly outperformed FwFM (2nd) (Table \ref{tab:rq3_p_values}), meaning that first-order features potentially improve effectiveness.

\begin{table}
    \centering
    \begin{tabular}{l|c}
        \hline
        Pair & p-value \\
        \hline
        NRM-F (2nd) - NRM-F (1st + 2nd) & 0.526 \\
        FwFM (2nd) - FwFM (1st + 2nd) & 0.0 \\
        \hline
    \end{tabular}
    \caption{P-values of paired t-tests on each method}
    \label{tab:rq3_p_values}
\end{table}

\section{Discussion}

The links between recommendation and information retrieval have also been explored elsewhere~\cite{Kallumadi2018}. This is one of the first works to link these two areas together in a practical sense under neural models. 

Contrary to our results, prior works have indicated that interaction-based models tend to be better than representation-based models. This comparison may not be fully appropriate because of two reasons. Firstly, it has been proved that feed forward neural networks model low-rank relations \cite{beutel2018latent}, meaning that simple implicit interaction learning models can mimic the behaviour of any explicit interaction learning models. Secondly, we showed that FwFM outperforms NRM-F, meaning that different architectures have different performance. The conclusion can vary depending on the model used for the experiment. A more detailed exploration on when interaction- vs representation-based models work seems to be needed.

The models that learned query-field interactions outperformed models trained using all interactions in both NRM-F and FwFM. This may be because query-field interactions are particularly important in document ranking, and adding irrelevant feature interactions may introduce noise into the model, resulting in degraded performance.

The original implementation of NRM-F does not use first-order field interactions. This may be because feeding those features are considered to overfit the model. For example, if there is a feature that indirectly represents the item's popularity, it could appear at the top of the list regardless of the user's intent. However, recommendation models usually do not care about the case. In a sense, feeding first-order features could improve performance because machine learning models eventually update parameters to minimize the loss.

In non-linear field interactions we showed how only FwFM's performance was improved. This may be because FwFM is more robust to noisy features to some extent since FwFM has weights that decrease the impact of not important features. The result suggests that feeding non linear interactions potentially improves performance.

\section{Conclusions}

We have shown how non-linearities in field-to-field interactions have an impact in ranking performance, potentially improving effectiveness.

Models that learned query-field interactions outperformed models trained using all interactions. Models including selected field-to-field features did not outperform models considering query-field interactions.

Our results also suggest an important effect of the chosen neural architecture on the performance of the ranking model, regardless of the type of interactions being considered.

\begin{acks}
TBD.
\end{acks}

\bibliographystyle{ACM-Reference-Format}
\bibliography{references}


\begin{thebibliography}{20}


\ifx \showCODEN    \undefined \def \showCODEN     #1{\unskip}     \fi
\ifx \showDOI      \undefined \def \showDOI       #1{#1}\fi
\ifx \showISBNx    \undefined \def \showISBNx     #1{\unskip}     \fi
\ifx \showISBNxiii \undefined \def \showISBNxiii  #1{\unskip}     \fi
\ifx \showISSN     \undefined \def \showISSN      #1{\unskip}     \fi
\ifx \showLCCN     \undefined \def \showLCCN      #1{\unskip}     \fi
\ifx \shownote     \undefined \def \shownote      #1{#1}          \fi
\ifx \showarticletitle \undefined \def \showarticletitle #1{#1}   \fi
\ifx \showURL      \undefined \def \showURL       {\relax}        \fi
\providecommand\bibfield[2]{#2}
\providecommand\bibinfo[2]{#2}
\providecommand\natexlab[1]{#1}
\providecommand\showeprint[2][]{arXiv:#2}

\bibitem[\protect\citeauthoryear{Beutel, Covington, Jain, Xu, Li, Gatto, and
  Chi}{Beutel et~al\mbox{.}}{2018}]%
        {beutel2018latent}
\bibfield{author}{\bibinfo{person}{Alex Beutel}, \bibinfo{person}{Paul
  Covington}, \bibinfo{person}{Sagar Jain}, \bibinfo{person}{Can Xu},
  \bibinfo{person}{Jia Li}, \bibinfo{person}{Vince Gatto}, {and}
  \bibinfo{person}{Ed~H Chi}.} \bibinfo{year}{2018}\natexlab{}.
\newblock \showarticletitle{Latent cross: Making use of context in recurrent
  recommender systems}. In \bibinfo{booktitle}{\emph{Proceedings of the
  Eleventh ACM International Conference on Web Search and Data Mining}}.
  \bibinfo{pages}{46--54}.
\newblock


\bibitem[\protect\citeauthoryear{Burges, Ragno, and Le}{Burges
  et~al\mbox{.}}{2007}]%
        {burges2007learning}
\bibfield{author}{\bibinfo{person}{Christopher~J Burges},
  \bibinfo{person}{Robert Ragno}, {and} \bibinfo{person}{Quoc~V Le}.}
  \bibinfo{year}{2007}\natexlab{}.
\newblock \showarticletitle{Learning to rank with nonsmooth cost functions}. In
  \bibinfo{booktitle}{\emph{Advances in neural information processing
  systems}}. \bibinfo{pages}{193--200}.
\newblock


\bibitem[\protect\citeauthoryear{Dehghani, Zamani, Severyn, Kamps, and
  Croft}{Dehghani et~al\mbox{.}}{2017}]%
        {Dehghani2017nrmweak}
\bibfield{author}{\bibinfo{person}{Mostafa Dehghani}, \bibinfo{person}{Hamed
  Zamani}, \bibinfo{person}{Aliaksei Severyn}, \bibinfo{person}{Jaap Kamps},
  {and} \bibinfo{person}{W.~Bruce Croft}.} \bibinfo{year}{2017}\natexlab{}.
\newblock \showarticletitle{Neural Ranking Models with Weak Supervision}. In
  \bibinfo{booktitle}{\emph{Proceedings of the 40th International ACM SIGIR
  Conference on Research and Development in Information Retrieval}} (Shinjuku,
  Tokyo, Japan) \emph{(\bibinfo{series}{SIGIR '17})}.
  \bibinfo{publisher}{Association for Computing Machinery},
  \bibinfo{address}{New York, NY, USA}, \bibinfo{pages}{65–74}.
\newblock
\showISBNx{9781450350228}
\urldef\tempurl%
\url{https://doi.org/10.1145/3077136.3080832}
\showDOI{\tempurl}


\bibitem[\protect\citeauthoryear{Huang, He, Gao, Deng, Acero, and Heck}{Huang
  et~al\mbox{.}}{2013}]%
        {huang2013learning}
\bibfield{author}{\bibinfo{person}{Po-Sen Huang}, \bibinfo{person}{Xiaodong
  He}, \bibinfo{person}{Jianfeng Gao}, \bibinfo{person}{Li Deng},
  \bibinfo{person}{Alex Acero}, {and} \bibinfo{person}{Larry Heck}.}
  \bibinfo{year}{2013}\natexlab{}.
\newblock \showarticletitle{Learning deep structured semantic models for web
  search using clickthrough data}. In \bibinfo{booktitle}{\emph{Proceedings of
  the 22nd ACM international conference on Information \& Knowledge
  Management}}. ACM, \bibinfo{pages}{2333--2338}.
\newblock


\bibitem[\protect\citeauthoryear{Kallumadi, Mitra, and Iofciu}{Kallumadi
  et~al\mbox{.}}{2018}]%
        {Kallumadi2018}
\bibfield{author}{\bibinfo{person}{Surya Kallumadi}, \bibinfo{person}{Bhaskar
  Mitra}, {and} \bibinfo{person}{Tereza Iofciu}.}
  \bibinfo{year}{2018}\natexlab{}.
\newblock \showarticletitle{A Line in the Sand: Recommendation or Ad-Hoc
  Retrieval? Overview of RecSys Challenge 2018 Submission by Team
  BachPropagate}. In \bibinfo{booktitle}{\emph{Proceedings of the ACM
  Recommender Systems Challenge 2018}} (Vancouver, BC, Canada)
  \emph{(\bibinfo{series}{RecSys Challenge '18})}.
  \bibinfo{publisher}{Association for Computing Machinery},
  \bibinfo{address}{New York, NY, USA}, Article \bibinfo{articleno}{7},
  \bibinfo{numpages}{6}~pages.
\newblock
\showISBNx{9781450365864}
\urldef\tempurl%
\url{https://doi.org/10.1145/3267471.3267478}
\showDOI{\tempurl}


\bibitem[\protect\citeauthoryear{Kim, Xue, and Croft}{Kim
  et~al\mbox{.}}{2009}]%
        {kim09prob}
\bibfield{author}{\bibinfo{person}{Jinyoung Kim}, \bibinfo{person}{Xiaobing
  Xue}, {and} \bibinfo{person}{W.~Bruce Croft}.}
  \bibinfo{year}{2009}\natexlab{}.
\newblock \showarticletitle{A Probabilistic Retrieval Model for Semistructured
  Data}. In \bibinfo{booktitle}{\emph{Advances in Information Retrieval}},
  \bibfield{editor}{\bibinfo{person}{Mohand Boughanem},
  \bibinfo{person}{Catherine Berrut}, \bibinfo{person}{Josiane Mothe}, {and}
  \bibinfo{person}{Chantal Soule-Dupuy}} (Eds.). \bibinfo{publisher}{Springer
  Berlin Heidelberg}, \bibinfo{address}{Berlin, Heidelberg},
  \bibinfo{pages}{228--239}.
\newblock
\showISBNx{978-3-642-00958-7}


\bibitem[\protect\citeauthoryear{Kim and Croft}{Kim and Croft}{2012}]%
        {kim2012field}
\bibfield{author}{\bibinfo{person}{Jin~Young Kim} {and}
  \bibinfo{person}{W~Bruce Croft}.} \bibinfo{year}{2012}\natexlab{}.
\newblock \showarticletitle{A field relevance model for structured document
  retrieval}. In \bibinfo{booktitle}{\emph{European Conference on Information
  Retrieval}}. Springer, \bibinfo{pages}{97--108}.
\newblock


\bibitem[\protect\citeauthoryear{Liu, Lu, Kurland, and Culpepper}{Liu
  et~al\mbox{.}}{2018}]%
        {Binsheng18field-based-relevance}
\bibfield{author}{\bibinfo{person}{Binsheng Liu}, \bibinfo{person}{Xiaolu Lu},
  \bibinfo{person}{Oren Kurland}, {and} \bibinfo{person}{J.~Shane Culpepper}.}
  \bibinfo{year}{2018}\natexlab{}.
\newblock \showarticletitle{Improving Search Effectiveness with Field-Based
  Relevance Modeling}. In \bibinfo{booktitle}{\emph{Proceedings of the 23rd
  Australasian Document Computing Symposium}} (Dunedin, New Zealand)
  \emph{(\bibinfo{series}{ADCS '18})}. \bibinfo{publisher}{Association for
  Computing Machinery}, \bibinfo{address}{New York, NY, USA}, Article
  \bibinfo{articleno}{11}, \bibinfo{numpages}{4}~pages.
\newblock
\showISBNx{9781450365499}
\urldef\tempurl%
\url{https://doi.org/10.1145/3291992.3292005}
\showDOI{\tempurl}


\bibitem[\protect\citeauthoryear{{Mitra} and {Craswell}}{{Mitra} and
  {Craswell}}{2018}]%
        {MitraandCraswellNIR18}
\bibfield{author}{\bibinfo{person}{B. {Mitra}} {and} \bibinfo{person}{N.
  {Craswell}}.} \bibinfo{year}{2018}\natexlab{}.
\newblock \bibinfo{booktitle}{}.
\newblock
\urldef\tempurl%
\url{https://doi.org/10.1561/1500000061}
\showDOI{\tempurl}


\bibitem[\protect\citeauthoryear{Mitra, Rosset, Hawking, Craswell, Diaz, and
  Yilmaz}{Mitra et~al\mbox{.}}{2019}]%
        {mitra2019incorporating}
\bibfield{author}{\bibinfo{person}{Bhaskar Mitra}, \bibinfo{person}{Corby
  Rosset}, \bibinfo{person}{David Hawking}, \bibinfo{person}{Nick Craswell},
  \bibinfo{person}{Fernando Diaz}, {and} \bibinfo{person}{Emine Yilmaz}.}
  \bibinfo{year}{2019}\natexlab{}.
\newblock \showarticletitle{Incorporating query term independence assumption
  for efficient retrieval and ranking using deep neural networks}.
\newblock \bibinfo{journal}{\emph{arXiv preprint arXiv:1907.03693}}
  (\bibinfo{year}{2019}).
\newblock


\bibitem[\protect\citeauthoryear{Nigam, Song, Mohan, Lakshman, Ding, Shingavi,
  Teo, Gu, and Yin}{Nigam et~al\mbox{.}}{2019}]%
        {DBLP:journals/corr/abs-1907-00937}
\bibfield{author}{\bibinfo{person}{Priyanka Nigam}, \bibinfo{person}{Yiwei
  Song}, \bibinfo{person}{Vijai Mohan}, \bibinfo{person}{Vihan Lakshman},
  \bibinfo{person}{Weitian Ding}, \bibinfo{person}{Ankit Shingavi},
  \bibinfo{person}{Choon~Hui Teo}, \bibinfo{person}{Hao Gu}, {and}
  \bibinfo{person}{Bing Yin}.} \bibinfo{year}{2019}\natexlab{}.
\newblock \showarticletitle{Semantic Product Search}.
\newblock \bibinfo{journal}{\emph{CoRR}}  \bibinfo{volume}{abs/1907.00937}
  (\bibinfo{year}{2019}).
\newblock
\showeprint[arxiv]{1907.00937}
\urldef\tempurl%
\url{http://arxiv.org/abs/1907.00937}
\showURL{%
\tempurl}


\bibitem[\protect\citeauthoryear{Ogilvie and Callan}{Ogilvie and
  Callan}{2003}]%
        {Ogilvie03known}
\bibfield{author}{\bibinfo{person}{Paul Ogilvie} {and} \bibinfo{person}{Jamie
  Callan}.} \bibinfo{year}{2003}\natexlab{}.
\newblock \showarticletitle{Combining Document Representations for Known-Item
  Search}. In \bibinfo{booktitle}{\emph{Proceedings of the 26th Annual
  International ACM SIGIR Conference on Research and Development in Informaion
  Retrieval}} (Toronto, Canada) \emph{(\bibinfo{series}{SIGIR '03})}.
  \bibinfo{publisher}{Association for Computing Machinery},
  \bibinfo{address}{New York, NY, USA}, \bibinfo{pages}{143–150}.
\newblock
\showISBNx{1581136463}
\urldef\tempurl%
\url{https://doi.org/10.1145/860435.860463}
\showDOI{\tempurl}


\bibitem[\protect\citeauthoryear{Pan, Xu, Ruiz, Zhao, Pan, Sun, and Lu}{Pan
  et~al\mbox{.}}{2018}]%
        {pan2018field}
\bibfield{author}{\bibinfo{person}{Junwei Pan}, \bibinfo{person}{Jian Xu},
  \bibinfo{person}{Alfonso~Lobos Ruiz}, \bibinfo{person}{Wenliang Zhao},
  \bibinfo{person}{Shengjun Pan}, \bibinfo{person}{Yu Sun}, {and}
  \bibinfo{person}{Quan Lu}.} \bibinfo{year}{2018}\natexlab{}.
\newblock \showarticletitle{Field-weighted factorization machines for
  click-through rate prediction in display advertising}. In
  \bibinfo{booktitle}{\emph{Proceedings of the 2018 World Wide Web
  Conference}}. \bibinfo{pages}{1349--1357}.
\newblock


\bibitem[\protect\citeauthoryear{Piwowarski and Gallinari}{Piwowarski and
  Gallinari}{2003}]%
        {piwowarski2003machine}
\bibfield{author}{\bibinfo{person}{Benjamin Piwowarski} {and}
  \bibinfo{person}{Patrick Gallinari}.} \bibinfo{year}{2003}\natexlab{}.
\newblock \showarticletitle{A machine learning model for information retrieval
  with structured documents}. In \bibinfo{booktitle}{\emph{International
  Workshop on Machine Learning and Data Mining in Pattern Recognition}}.
  Springer, \bibinfo{pages}{425--438}.
\newblock


\bibitem[\protect\citeauthoryear{Robertson, Walker, Jones, Hancock-Beaulieu,
  and Gatford}{Robertson et~al\mbox{.}}{1995}]%
        {robertson1995okapi}
\bibfield{author}{\bibinfo{person}{Stephen Robertson}, \bibinfo{person}{S.
  Walker}, \bibinfo{person}{S. Jones}, \bibinfo{person}{M.~M.
  Hancock-Beaulieu}, {and} \bibinfo{person}{M. Gatford}.}
  \bibinfo{year}{1995}\natexlab{}.
\newblock \showarticletitle{Okapi at TREC-3}. In
  \bibinfo{booktitle}{\emph{Overview of the Third Text REtrieval Conference
  (TREC-3)} (\bibinfo{edition}{overview of the third text retrieval conference
  (trec–3)} ed.)}. \bibinfo{publisher}{Gaithersburg, MD: NIST},
  \bibinfo{pages}{109--126}.
\newblock
\urldef\tempurl%
\url{https://www.microsoft.com/en-us/research/publication/okapi-at-trec-3/}
\showURL{%
\tempurl}


\bibitem[\protect\citeauthoryear{Robertson, Zaragoza, and Taylor}{Robertson
  et~al\mbox{.}}{2004}]%
        {Robertson2004}
\bibfield{author}{\bibinfo{person}{Stephen Robertson}, \bibinfo{person}{Hugo
  Zaragoza}, {and} \bibinfo{person}{Michael Taylor}.}
  \bibinfo{year}{2004}\natexlab{}.
\newblock \showarticletitle{Simple BM25 Extension to Multiple Weighted Fields}.
  In \bibinfo{booktitle}{\emph{Proceedings of the Thirteenth ACM International
  Conference on Information and Knowledge Management}} (Washington, D.C., USA)
  \emph{(\bibinfo{series}{CIKM '04})}. \bibinfo{publisher}{Association for
  Computing Machinery}, \bibinfo{address}{New York, NY, USA},
  \bibinfo{pages}{42–49}.
\newblock
\showISBNx{1581138741}
\urldef\tempurl%
\url{https://doi.org/10.1145/1031171.1031181}
\showDOI{\tempurl}


\bibitem[\protect\citeauthoryear{Sarvi, Voskarides, Mooiman, Schelter, and
  de~Rijke}{Sarvi et~al\mbox{.}}{2020}]%
        {sarvi2020comparison}
\bibfield{author}{\bibinfo{person}{Fatemeh Sarvi}, \bibinfo{person}{Nikos
  Voskarides}, \bibinfo{person}{Lois Mooiman}, \bibinfo{person}{Sebastian
  Schelter}, {and} \bibinfo{person}{Maarten de Rijke}.}
  \bibinfo{year}{2020}\natexlab{}.
\newblock \showarticletitle{A Comparison of Supervised Learning to Match
  Methods for Product Search}.
\newblock \bibinfo{journal}{\emph{arXiv preprint arXiv:2007.10296}}
  (\bibinfo{year}{2020}).
\newblock


\bibitem[\protect\citeauthoryear{Svore and Burges}{Svore and Burges}{2009}]%
        {svore2009machine}
\bibfield{author}{\bibinfo{person}{Krysta~M Svore} {and}
  \bibinfo{person}{Christopher~JC Burges}.} \bibinfo{year}{2009}\natexlab{}.
\newblock \showarticletitle{A machine learning approach for improved BM25
  retrieval}. In \bibinfo{booktitle}{\emph{Proceedings of the 18th ACM
  conference on Information and knowledge management}}.
  \bibinfo{pages}{1811--1814}.
\newblock


\bibitem[\protect\citeauthoryear{Wilkinson}{Wilkinson}{1994}]%
        {wilkinson1994effective}
\bibfield{author}{\bibinfo{person}{Ross Wilkinson}.}
  \bibinfo{year}{1994}\natexlab{}.
\newblock \showarticletitle{Effective retrieval of structured documents}. In
  \bibinfo{booktitle}{\emph{SIGIR’94}}. Springer, \bibinfo{pages}{311--317}.
\newblock


\bibitem[\protect\citeauthoryear{Zamani, Mitra, Song, Craswell, and
  Tiwary}{Zamani et~al\mbox{.}}{2017}]%
        {DBLP:journals/corr/abs-1711-09174}
\bibfield{author}{\bibinfo{person}{Hamed Zamani}, \bibinfo{person}{Bhaskar
  Mitra}, \bibinfo{person}{Xia Song}, \bibinfo{person}{Nick Craswell}, {and}
  \bibinfo{person}{Saurabh Tiwary}.} \bibinfo{year}{2017}\natexlab{}.
\newblock \showarticletitle{Neural Ranking Models with Multiple Document
  Fields}.
\newblock \bibinfo{journal}{\emph{CoRR}}  \bibinfo{volume}{abs/1711.09174}
  (\bibinfo{year}{2017}).
\newblock
\showeprint[arxiv]{1711.09174}
\urldef\tempurl%
\url{http://arxiv.org/abs/1711.09174}
\showURL{%
\tempurl}


\end{thebibliography}

\end{document}